\documentclass[12pt]{article}  
\usepackage{amsmath,amssymb,amsfonts}
\usepackage{longtable}

\def\mylist{\begin{list}{}{\setlength{\leftmargin}{0.5in}
               \setlength{\listparindent}{-0.5in}
               \setlength{\itemindent}{\listparindent}}}

\newcommand{\nD}[1]{\not D}
\newcommand{\cN}{{\mathcal N}}
\newcommand{\RR}{{\mathbb R}}

\newcommand{\ZZ}{{\mathbb Z}}

\newcommand{\Tr}{{\rm Tr}}

\newcommand{\sign}{{\rm sign}}

\begin{document}

\begin{titlepage}

\title{Seiberg-like duality in three dimensions for orthogonal gauge groups}
\author{Anton Kapustin\\ {\it California Institute of Technology}}

\maketitle

\abstract{We propose a duality for $\cN=2$ $d=3$ Chern-Simons gauge theories with orthogonal gauge groups and matter in the vector representation. This duality generalizes level-rank duality for pure Chern-Simons gauge theories with orthogonal gauge groups and is reminiscent of Seiberg duality in four dimensions. We perform extensive checks by comparing partition functions of theories related by dualities. We also determine the conformal dimensions of fields using Z-extremization.}

\end{titlepage}

\section{Introduction and summary}

After N. Seiberg's seminal paper \cite{seiberg}, Seiberg duality for $\cN=1$ $d=4$ gauge theories has been the subject of many works, but it is still insufficiently understood. Recently, some new dualities for $\cN=2$ $d=3$ Chern-Simons gauge theories reminiscent of Seiberg duality have been proposed and studied \cite{GK,KWY3,WY}.There are several reasons to be interested in these dualities. One reason is the connection with the level-rank duality of topological Chern-Simons gauge theories. Unlike in four dimensions, the 3d duality makes sense even with no matter fields at all, if Chern-Simons coupling is large enough, and reduces in this case to the level-rank duality \cite{GK,KWY3}. Since the latter is rather well understood, one can try to apply the insights from the level-rank duality to 3d Seiberg-like dualities. Another reason is that in 3d we know how to compute the expectation values of Wilson loop operators exactly \cite{KWY1}. This could help to determine how Wilson loops map under duality, something which remains a mystery in four dimensions. 

In four dimensions there are three infinite series of Seiberg dualities, as well as several exceptional ones. The three infinite series correspond to gauge groups $SU(N)$, $USp(2N)$ and $SO(N)$ and matter in the fundamental representation. Let $N_f$ determine the number of matter flavors.\footnote{A flavor in the $SU(N)$ case consists of a pair of chiral superfields in representations ${\bf N}$ and $\bar {\bf N}$. In the $SO(N)$ case a flavor is a single chiral superfield  in the representation ${\bf N}$. In the $USp(2N)$ case a flavor is a pair of chiral superfields in the representation ${\bf 2N}$. Note that in the latter case the number of chiral superfields is even to avoid a global gauge anomaly \cite{Wittenglobal}.} Seiberg duality maps gauge groups in each series according to the following rules \cite{seiberg,IP}:
\begin{eqnarray}
SU(N_c) &\mapsto & SU(N_f-N_c),\\
USp(2N_c) & \mapsto & USp(2N_f-2N_c-4),\\
SO(N_c) & \mapsto & SO(N_f-N_c+4).
\end{eqnarray}
The dual theory (known as ``magnetic'' theory) also has some chiral superfields which are gauge singlets and couple to dual quarks via a cubic superpotential.

Giveon and Kutasov \cite{GK} proposed that the first of these series has a analog for $\cN=2$ $d=3$ Chern-Simons gauge theories with unitary groups. The rule for mapping of 3d gauge groups is
\begin{equation}\label{unitary3d}
U(N_c)_k\mapsto U(N_f-N_c+|k|)_{-k},
\end{equation}
where a subscript on a gauge group denotes the Chern-Simons coupling. The singlets in the ``magnetic'' theory and the ``magnetic'' superpotential are the same as in 4d. This duality has been extensively tested in \cite{KWY3,WY} using supersymmetric localization. In particular, it has been proved by Willett and Yaakov \cite{WY} that the $S^3$ partition functions for the dual theories agree. 

The symplectic Seiberg duality also has a 3d analog. The gauge groups map according to the rule
\begin{equation}\label{symplectic3d}
USp(2N_c)_k\mapsto USp(2(N_f-N_c-1+|k|))_{-k},
\end{equation}
and the superpotential in the``magnetic'' theory is the same as in the 4d case.
This duality becomes rather natural if we recall that there are two other conjectural dualities for $\cN=2$ $d=3$ gauge theories  proposed by Aharony \cite{Aharony}. Aharony's dualities involve gauge theories without Chern-Simons terms. Thus the theories are not conformal on the classical level and flow to strongly-coupled IR fixed points. One of Aharony's dualities maps ``electric'' $U(N_c)$ gauge theory with $N_f$ flavors to ``magnetic'' $U(N_f-N_c)$  gauge theory with $N_f$ flavors, some singlet fields, and a superpotential. The other duality maps $USp(2N_c)$ gauge theory with $N_f$ flavors to $USp(2N_f-2N_c-2)$ gauge theory with $N_f$ flavors, some singlet fields, and a superpotential. In the absence of Chern-Simons terms the singlets fields are somewhat different than in 4d, and the superpotential contains new terms coupling singlets to monopole operators \cite{Aharony}. All these complicating features disappear if we give several flavors a large real mass and integrate them out. This generates a Chern-Simons term both in the ``electric'' and ``magnetic'' theories, modifies the mapping of gauge groups to incorporate Chern-Simons level, and on the ``magnetic'' side removes the extra singlets which  couple to monopole operators. In the unitary case, as noted in \cite{GK}, this gives Seiberg-like duality with Chern-Simons terms (\ref{unitary3d}). The logic applies equally well in the symplectic case and gives Seiberg-like duality with Chern-Simons terms (\ref{symplectic3d}). The symplectic duality has been proved to hold on the level of partition functions for all $N_f,N_c$ and $k$ in \cite{WY}. 

In this paper we propose and test a 3d analog of orthogonal Seiberg duality. The proposed mapping of gauge groups is
$$
O(N_c)_k\mapsto O(N_f-N_c+|k|+2)_{-k}.
$$
The ``electric'' $O(N_c)$ gauge theory has $N_f$ flavors of chiral superfields $Q^i$, $i=1,\ldots,N_f$ in the vector representation and no superpotential.  The ``magnetic'' $O(N_f-N_c+|k|+2)$ gauge theory has $N_f$ flavors of chiral superfields $q_i$ in the vector representation as well as a singlet chiral superfield $M^{ij}$ which is a symmetric $N_f\times N_f$ matrix. The superpotential in the ``magnetic'' theory is 
$$
W=q_i q_j M^{ij}.
$$
We perform extensive checks of our proposal using supersymmetric localization. We use Z-extremization to find the conformal dimensions of the fields. We also outline an $\cN=3$ version of these dualities and their relationship with orthogonal level-rank duality.

Note that dual gauge group is not the same as in the 4d case even if we formally set $k=0$ (this also applies to 3d symplectic duality). It will also prove important that the duality involves orthogonal rather than special orthogonal groups.

I would like to thank Denis Bashkirov for discussions and the Simons Center for Geometry and Physics for hospitality. This work was supported in part by the DOE grant DE-FG02-92ER40701.

\section{Orthogonal level-rank duality and its generalizations}

To motivate our proposal, let us recall how level-rank duality works for orthogonal Chern-Simons gauge theories \cite{levelrank1,levelrank2}. As usually formulated, orthogonal level-rank duality relates topological Chern-Simons theories with gauge groups $SO(N_c)_k$ and $SO(|k|)_{-N_c}$. Here Chern-Simons level $k$ is the ``bare'' Chern-Simons level. Quantum effects lead to a renormalization $k\mapsto K=k+(N_c-2)\sign\, k$, since the dual Coxeter number of $SO(N_c)$ is $N_c-2$. If we use the renormalized level $K$ to label Chern-Simons theories, level-rank duality becomes
$$
O(N_c)_K\mapsto O(|K|+2-N_c)_{-K}
$$
Here we replaced special orthogonal groups with orthogonal groups in a somewhat ad hoc manner. While many discussions of level-rank duality are insensitive to this difference, below we will see strong evidence in favor of dropping the special condition. This is also more natural from the 2d point of view where orthogonal level-rank duality appears from considering symmetries of a system of $k N_c$ Mayorana-Weyl fermions.

Bosonic Chern-Simons gauge theory with renormalized level $K$ is equivalent to $\cN=3$ Chern-Simons gauge theory with no matter fields and Chern-Simons coupling 
 $K$. From now one we will always use the $\cN=3$ normalization of the Chern-Simons coupling and will denote it by $k$ rather than $K$. We can try to generalize level-rank duality while preserving $\cN=3$ supersymmetry by introducing $L$ hypermultiplets in the vector representation. Such a hypermultiplet consists of a pair of chiral superfields in the vector representation, so from the $\cN=2$ viewpoint we are introducing $2L$ flavors of chiral superfields. Altogether, we can describe matter superfields in the ``electric'' theory by a superfield matrix $Q^i_A$ of size $2L\times N_c$. $\cN=3$ supersymmetry requires the superpotential to be
$$
W=\sqrt 2\Tr (Q^t JQ\Phi)+\frac{k}{4\pi}\Tr\,\Phi^2.
$$
Here $\Phi$ is a superfield in the adjoint of the gauge group, i.e. a 2nd rank anti-symmetric tensor $\Phi_{AB}$, and $J=J_{ij}$ is a non-degenerate anti-symmetric tensor of rank $2L$ in flavor space. This superpotential preserves $USp(2L)$ flavor symmetry. A natural guess for the mapping of gauge groups in the $\cN=3$ case is then
\begin{equation}\label{orthogonal3dK}
O(N_c)_K\mapsto O(N_f+|K|+2-N_c)_{-K}
\end{equation}
The dual theory is also an $\cN=3$ theory with $L=N_f/2$ hypermultiplets in the vector representation.

The superfield $\Phi$ is auxiliary (has no kinetic term other than the one coming from the superpotential). Integrating it out we get a quartic superpotential for $Q$:
$$
W=-\frac{2\pi}{k}\Tr (Q^t JQ)^2.
$$
Thus we can reinterpret the above $\cN=3$ duality as the duality of $\cN=2$ theories with $2L$ chiral superfields in the vector representation and a quartic superpotential.

Next we would like to generalize this to the $\cN=2$ case. We keep the mapping of gauge groups unchanged, but on the ``electric'' side set the superpotential to zero. Unlike in the $\cN=3$ case there is no need to take $N_f$ to be even, so we allow it to be an arbitrary whole number. The flavor symmetry group is enlarged to $U(N_f)$, and matter superfields $Q^i, i=1,\ldots,N_f$ transform in the representation ${\bf N_f}$ of $U(N_f)$. On the ``magnetic'' side by analogy with the 4d case we introduce a singlet chiral superfield $M^{ij}$ which is a symmetric $N_f\times N_f$ matrix and postulate a superpotential 
$$
W=\sqrt 2 q_i q_j M^{ij}=\sqrt 2 q M q^t.
$$
The ``magnetic'' superfields $q_i$ transform in the representation $\bar {\bf N}$ of $U(N_f)$ and vector representation of the ``magnetic'' gauge group. 

The flavor symmetry of both ``electric'' and `magnetic'' theories is $U(N_f)$. On the ``electric'' side there is also $U(1)_R$ symmetry with respect to which superfields $Q^i$ have charge $0$.  On the ``magnetic'' side the corresponding $U(1)_R$ symmetry acts trivially on $M$, while superfields $q_i$ have charge $1$ with respect to it.

If either $N_c$ or $N_c'$ are equal to $2$, there is also a conserved topological current (the dual of the gauge field strength). However, this current changes sign under gauge transformations which are in the component of $O(2)$ disconnected from the identity element. In other words, the difference between the $O(2)$ gauge group and the $SO(2)=U(1)$ gauge group is that in the former case charge conjugation symmetry is gauged. Charge conjugation flips the sign of the topological current, therefore it is not gauge-invariant in the $O(2)$ case. This is important, since our duality typically maps $O(2)$ gauge theory to a nonabelian gauge theory which does not admit a conserved topological current.

Next let us consider the mapping of gauge-invariant chiral primary operators. On the ``electric'' side we have a composite meson $Q^i_A Q^j_B\delta^{AB}=Q^i Q^j$ which is a symmetric $N_f\times N_f$ matrix. On the ``magnetic'' side the corresponding field is the singlet superfield $M^{ij}$. The ``magnetic'' composite meson $q_i q_j$ is not primary thanks to the superpotential in the ``magnetic'' theory. Unlike in the 4d case, there are no baryon operators, because they are not invariant under the gauge group. Here the distinction between $SO(N)$ and $O(N)$ is again important. There are no monopole operators which are chiral primaries thanks to the Chern-Simons term. 

The relationship between $\cN=2$ and $\cN=3$ dualities  is the following. Suppose $N_f$ is even, $N_f=2L$. Let us perturb the ``electric'' side with a quartic superpotential 
$$
W_{el}=-\frac{2\pi}{k}\Tr (Q^t J Q)^2,
$$
where the trace is over gauge indices and $J_{ij}$ is a symplectic form in $2L$-dimensional flavor space. This perturbation drives the theory to an $\cN=3$ fixed point.
If we identify $QQ^t=\lambda M$ where $\lambda$ is an unknown constant, then the corresponding superpotential on the ``magnetic'' side is
$$
W_{mag}=\sqrt 2 q M q^t-\frac{2\pi}{k}\lambda^2\Tr (JM)^2.
$$
Assuming that in the infrared the D-term for $M$ is irrelevant, we may treat $M$ as an auxiliary superfield and integrate it out. This gives the correct quartic superpotential on the ``magnetic'' side:
$$
W_{mag}=\frac{k}{4\pi\lambda^2} \Tr (qJ q^t)^2.
$$
Thus $\cN=2$ duality implies $\cN=3$ duality provided  $\lambda=\pm k/2\pi\sqrt 2$.

\section{Partition function on a 3-sphere}

We can test $\cN=2$ dualities by computing the partition function on a 3-sphere using supersymmetric localization. Originally localization was applied to theories with at least $\cN=3$ supersymmetry \cite{KWY1}, but it was pointed out in \cite{Jafferis}  that the method can be generalized to arbitrary $\cN=2$ theories. The main difference compared to the $\cN=3$ case is that conformal dimensions of fields may get quantum corrections. It was argued in \cite{Jafferis} that quantum-corrected dimensions can be inferred by treating them as free parameters and extremizing the absolute value of the partition function with respect to them.

In fact, partition functions of dual theories must agree not only for the critical value of conformal dimensions, but for all values. This follows from the arguments in \cite{Jafferis} where the choice of conformal dimension was shown to control the choice of a supercharge used for localization and the curvature couplings in the $S^3$ action. The critical value simply corresponds to the special case where the supersymmetric theory on $S^3$ is equivalent to a superconformal theory on $\RR^3$. 

Keeping this in mind, we conclude that the $S^3$ partition functions of dual theories must agree as functions of conformal dimensions. For simplicity, on the ``electric'' side we only consider choices of conformal dimensions which preserve the full flavor symmetry. Then there is only one free parameter, the conformal dimension $\Delta$ of the composite meson $Q^i Q^j$. Unitarity requires the critical value of $\Delta$ (the one which extremizes $|Z|$) to be greater or equal than $1/2$. It is a nontrivial consistency check that in all cases we studied the critical value satisfies this constraint. 

On the ``magnetic'' side we have to assign dimension $\Delta$ to the singlet superfield $M^{ij}$, therefore ``magnetic''  superfields $q_i$ must have dimension $1-\Delta/2$, to ensure that the superpotential has dimension $2$. 

Supersymmetric localization gives the following expression for the partition function of the ``electric'' theory \cite{KWY1,Jafferis}:
\begin{eqnarray}
Z^{el,N_f}_{N_c,k}(\Delta)=\frac{1}{|\mathcal W|}\int  \left(\prod_a du^a\right) F_{N_c}(u) e^{N_f G_{N_c}(u,\Delta)+k i \pi \sum_b u_b^2},
\end{eqnarray}
where the variables $u^a$ are real, the indices $a,b$ range from $1$ to $\left[N_c/2\right]$, and $|\mathcal W|$ is the order of the Weyl group $\mathcal W$. 
The expressions for functions $F_{N_c}$ and $G_{N_c}$ depend on whether $N_c$ is even or odd. If $N_c$ is even, $N_c=2n$, we have
\begin{eqnarray}
F_{2n}(u)&=&\prod_{a<b} \left(4 \sinh(\pi (u_a-u_b))\sinh (\pi(u_a+u_b))\right)^2,\\
 G_{2n}(u,\Delta)&=&\sum_a \left(\ell(1-\Delta/2+i u_a)+\ell(1-\Delta/2-i u_a)\right).
\end{eqnarray}
Here the function $\ell(z)$ is given by 
$$
\ell(z)=-z {\rm log} (1-e^{2\pi iz})+\frac{i}{2} \left(\pi z^2+\frac{1}{\pi}{\rm Li}_2(e^{2\pi i z})\right)-\frac{i\pi}{12}.
$$
If $N_c$ is odd, $N_c=2n+1$, we have
\begin{eqnarray}
F_{2n+1}(u)&=&\prod_c (2\sinh(\pi u_c))^2\prod_{a<b} \left(4 \sinh(\pi (u_a-u_b))\sinh (\pi(u_a+u_b))\right)^2,\\
 G_{2n+1}(u,\Delta)&=&\ell(1-\Delta/2)+\sum_a \left(\ell(1-\Delta/2+i u_a)+\ell(1-\Delta/2-i u_a)\right).
\end{eqnarray}

The partition function of the ``magnetic'' theory is similar:
$$
Z^{mag,N_f}_{N_c,k}(\Delta)=\frac{1}{|\mathcal W'|}e^{\ell(1-\Delta)N_f(N_f+1)/2} \int \left(\prod_a du^a\right) F_{N_c'}(u) e^{N_f G_{N_c'}(u,2-\Delta)-k i \pi \sum_b u_b^2},
$$
where $N_c'=N_f-N_c+|k|+2$ and the indices $a,b$ now range from $1$ to $[N_c'/2]$. The pre-factor is the contribution of $N_f(N_f+1)/2$ singlet chiral superfields of dimension $\Delta$.

We computed the partition functions numerically for all possible dual pairs in the range $1\leq N_c,N_c'\leq 4$ and verified that the values of ${\rm log} |Z|$ agree with accuracy of at least $10^{-5}$. We did not try to match the phase because it is affected by framing ambiguities. Let us discuss a few examples. We will denote by  $O(N_c)^{N_f}_k$ a theory with gauge group $O(N_c)$, $N_f$ flavors of matter fields, and Chern-Simons level $k$. On the ``magnetic'' side, we also need to add singlet superfields $M^{ij}$.

Our first example is electric $O(4)^1_1$ theory. That is, the gauge group is $O(4)$, with Chern-Simons coupling $1$ and a single chiral superfield $Q$ in the vector representation. The ``magnetic'' gauge group is $O(0)$, i.e. trivial. The ``magnetic'' theory has a single chiral superfield $M$ and no superpotential. Thus the ``magnetic'' theory is free. Duality predicts that the dimension of the composite meson $Q^2$ in the electric theory is $1/2$. The values of partition functions agree for all values of $\Delta$ and are tabulated in Table~\ref{table0}. 
Extremization of the partition function of a free chiral superfield gives $\Delta_0=1/2$ \cite{Jafferis}. Since the partition functions of the two theories agree to an accuracy of at least $10^{-5}$, numerical extremization of the partition function of the electric theory gives the same result with a very good accuracy.

\begin{table}\caption{Free energy $F=-{\rm log} |Z|$ for electric $O(4)^1_1$ and a free theory of a single chiral superfield}\label{table0}
\begin{center}
\begin{tabular}{|c|ccccccc|}\hline
$\Delta$ &  0.35 & 0.4 & 0.45 & 0.50 & 0.55 & 0.60 & 0.65 \\ \hline
$-{\rm log} |Z|$ & 0.27724 & 0.31813 & 0.33997 & 0.34657  & 0.32484 & 0.30050 & 0.32484  \\ \hline
\end{tabular}
\end{center}

\end{table}

Next consider electric $O(3)^1_1$ theory. Its magnetic dual $O(1)^1_{-1}$ has a discrete gauge group $O(1)=\ZZ_2$ and can be described as a Wess-Zumino model with two chiral superfields $q$ and $M$ and a superpotential
$$
W=M q^2.
$$
This WZ model has a one-dimensional moduli space parameterized by $M$. This agrees with the moduli space of the ``electric'' theory which is parametrized by $Q^2$. The chiral rings also agree provided we take into account discrete $O(1)$ symmetry. On the ``electric'' side the chiral ring is generated by $Q^2$, with no relations. On the ``magnetic'' side the only gauge-invariant chiral primary is $M$. The field $q$ is primary, but is odd under $O(1)=\ZZ_2$. The field $q^2$ is $O(1)$-invariant, but is not  primary thanks to the superpotential. The partition functions are tabulated in Table \ref{table1}. 
\begin{table}\caption{Free energy $F=-{\rm log} |Z|$ for electric $O(3)^1_1$ and magnetic $O(1)^1_{-1}$ theories}\label{table1}
\begin{center}
\begin{tabular}{|c|cccccc|}\hline
$\Delta$ &  0.5 & 0.6 & 0.7 & 0.8 & 0.9 & 1.0 \\ \hline
$-{\rm log} |Z|$ & 1.27214 & 1.28726 & 1.26292 & 1.20907 & 1.13284 & 1.03972\\ \hline
\end{tabular}
\end{center}

\end{table}

The conformal dimension $\Delta_0$ can be obtained numerically by extremizing ${\rm log} |Z|$. Since the ``magnetic'' theory is so simple in this case, we can perform extremization analytically and find that $\Delta_0$ is the root of the equation
$$
2\left(\tan^2\frac{\pi\Delta}{2}-1\right)=\frac{\Delta}{1-\Delta}.
$$
There is indeed a unique root in the physical range $\Delta_0\geq 1/2$ which is given numerically by 
$$
\Delta_0=0.58353\ldots
$$
One can show that  $\Delta_0$ is irrational. $\Delta_0$ is the conformal dimension of $M$; the conformal dimension of $q$ is $1-\Delta_0/2$. Note that the ``magnetic'' theory is parity-invariant, while the ``electric'' one is not parity-invariant on the classical level. Thus duality predicts that the ``electric'' theory has hidden parity-invariance.

Next consider ``electric'' $O(3)^1_2$ theory. The ``magnetic'' theory is $O(2)^1_{-2}$ and is an abelian gauge theory. The values of the partition function are tabulated in Table \ref{table2}. The critical value of the dimension is $\Delta_0=0.71186\ldots$ in both theories. Note that the dimension $\Delta_0$ of the composite meson is now closer to the classical value $1$. This is the expected result: as $k$ increases, the theory becomes more weakly coupled.

\begin{table}\caption{Free energy $F=-{\rm log} |Z|$ for electric $O(3)^1_2$ and magnetic $O(2)^1_{-2}$ theories}\label{table2}
\begin{center}
\begin{tabular}{|c|cccccc|}\hline
$\Delta$ &  0.5 & 0.6 & 0.7 & 0.8 & 0.9 & 1.0 \\ \hline
$-{\rm log} |Z|$ & 1.88474 & 1.94875 & 1.97040 & 1.95897 & 1.92093 & 1.86112 \\ \hline
\end{tabular}
\end{center}

\end{table}

\begin{table}\caption{Free energy $F=-{\rm log} |Z|$ for electric $O(3)^1_3$ and magnetic $O(2)^1_{-3}$ theories}\label{table3}
\begin{center}
\begin{tabular}{|c|cccccc|}\hline
$\Delta$ &  0.5 & 0.6 & 0.7 & 0.8 & 0.9 & 1.0 \\ \hline
$-{\rm log} |Z|$ & 2.27080 & 2.36744 & 2.41926 & 2.43522 & 2.42151 & 2.38276 \\ \hline
\end{tabular}
\end{center}

\end{table}

\begin{table}\caption{Free energy $F=-{\rm log} |Z|$ for electric $O(2)^2_2$ and magnetic $O(4)^2_{-2}$ theories}\label{table4}
\begin{center}
\begin{tabular}{|c|cccccc|}\hline
$\Delta$ &  0.5 & 0.6 & 0.7 & 0.8 & 0.9 & 1.0 \\ \hline
$-{\rm log} |Z|$ & 2.25388 & 2.50772 & 2.65955 & 2.73376 & 2.74681 & 2.71058 \\ \hline
\end{tabular}
\end{center}

\end{table}

If we take ``electric'' $O(3)^1_3$ theory, the magnetic theory is $O(3)^1_3$. This is the ``self-dual'' case, in the sense that the ``magnetic'' gauge group is the same as the ``electric'' one. The partition functions are tabulated in Table \ref{table3}. The conformal dimension $\Delta_0=0.80085...$ is even closer to $1$.

Finally, let us give another example of duality involving a rank-two gauge group. Consider electric $O(2)^2_2$ theory. The magnetic theory is $O(4)^2_{-2}$. The partition function is tabulated in Table \ref{table4}. The conformal dimension is $\Delta_0\simeq 0.87364$ in both cases. 

For other theories we have considered we list quantum conformal dimensions $\Delta_0$ in Table \ref{table5}. We only list ``electric'' gauge theories, but we verified that the dimension is the same for ``magnetic'' theories.

These computations provide strong evidence in favor of Seiberg-like duality in three dimensions for orthogonal gauge groups and vector matter. It would be interesting to find an analytic proof of the duality on the level of partition functions, along the lines of \cite{WY}. It would also be interesting to extend the duality to theories involving the matter in both vector and spinor representations of the gauge group.

\begin{table}\caption{Quantum conformal dimensions for various ``electric'' theories}\label{table5}

\begin{center}
\begin{tabular}{|c|ccccccc|} \hline
{\rm Theory} & $O(2)^1_1$ & $O(2)^1_2$ & $O(2)^2_1$ & $O(3)^2_1$ & $O(3)^2_2$ & $O(3)^3_2$ & $O(3)^2_3$ \\ \hline
$\Delta_0$  & 0.76893 & 0.85121 & 0.83965 & 0.66649 & 0.74917 & 0.78675 & 0.80743\\ \hline
\end{tabular}
\end{center}
\end{table}

\end{document}